\documentclass[12pt]{article}

\catcode`\@=11
\@addtoreset{equation}{section}

\global\arraycolsep=2pt
\usepackage{mathrsfs,amsbsy,amssymb,latexsym,amsfonts,amsmath,cite}

\newcommand\CL{{\mathcal L}}

\newcommand\CF{{\mathcal F}}
\newcommand\CM{{\mathcal M}}
\newcommand\CH{{\mathcal H}}
\newcommand\CI{{\mathcal I}}
\newcommand\CJ{{\mathcal J}}
\newcommand\CC{{\mathcal C}}
\newcommand\CB{{\mathcal B}}

\newcommand\SC{{\mathscr{C}}}

\newcommand\BL{{\bf{L}}}

\renewcommand\mod{~\text{mod}~}

\newcommand\DBI{{\mathrm{DBI}}}
\renewcommand\NG{{\mathrm{NG}}}
\newcommand\WZ{{\mathrm{WZ}}}

\newcommand\e{\mathrm{e}}
\newcommand\diag{\mathrm{diag}}
\newcommand\vol{\mathrm{vol}}

\newcommand{\adss}[2]{{AdS$_{#1}\times$S$^{#2}$}}
\newcommand{\ads}[1]{{AdS$_{#1}$}}
\newcommand{\s}[1]{{S$^{#1}$}}

\newcommand\Mu{{\underline{M}}}
\newcommand\Mb{{\bar M}}

\begin{document}

\begin{flushright}
\parbox{4.2cm}
{{\tt hep-th/0703061} \\ 
KEK-TH-1125 \hfill \\ 
OIQP-06-21 \hfill 
}
\end{flushright}

\vspace*{0.5cm}

\begin{center}
{\Large \bf
Non-relativistic String and D-branes on \adss{5}{5} \\
from Semiclassical Approximation
}
\end{center}
\vspace{10mm}

\centerline{\large Makoto Sakaguchi$^{a}$
 and Kentaroh Yoshida$^{b}$
}

\vspace{8mm}

\begin{center}
$^a$ 
{\it Okayama Institute for Quantum Physics\\
1-9-1 Kyoyama, Okayama 700-0015, Japan} \\
{\tt makoto\_sakaguchi\_at\_pref.okayama.jp}
\vspace{5mm}

$^b${\it Theory Division, Institute of Particle and Nuclear Studies, \\ 
High Energy Accelerator Research 
Organization (KEK),\\ 
1-1 Oho, Tsukuba, Ibaraki 305-0801, Japan.} 
\\
{\tt kyoshida\_at\_post.kek.jp}
\end{center}

\vspace{1cm}

\begin{abstract} 
We show that non-relativistic actions of string and D-branes on 
\adss{5}{5}
can be reproduced as a semiclassical approximation of
the 
string and D-brane actions
around static 1/2 BPS configurations. This
is contrastive to the Penrose limit. For example,
the pp-wave string can be recaptured as a semiclassical approximation
around a 1/2 BPS particle rotating at the velocity of light. We argue
that  small deformation of a straight Wilson line would give a composite
operator in the gauge-theory side that corresponds to a semiclassical
state of the non-relativistic string, according to the semiclassical
interpretation.

\end{abstract}

\thispagestyle{empty}
\setcounter{page}{0}

\newpage
\section{Introduction}

AdS/CFT duality \cite{AdS/CFT:M,AdS/CFT:GKPW} has now been widely accepted and
continues to be an important laboratory to look for new aspects of gauge
theories as well as string theory. The duality however has not been
rigorously proved and it is obviously important to find a further support
and confirm it furthermore. One of difficulties is to solve type IIB
string on \adss{5}{5} definitely. The Green-Schwarz action constructed by
Metsaev and Tseytlin \cite{MT} is too complicated to quantize it. Although
it is suspected to be quantum integrable since Bena, Polchinski and Roiban
have pointed out the classical integrability of AdS superstring
\cite{BPR}, the attempts for the quantization have not been succeeded yet. 
Thus
it would still be a nice direction to seek a soluble subsector.

\medskip 

Indeed, Berenstein, Maldacena and Nastase found a soluble sector of AdS/CFT
\cite{BMN} by using the Penrose limit \cite{Penrose}. The action of AdS
superstring is reduced to the pp-wave string, which becomes a free
massive theory on the flat world-sheet by taking a light-cone gauge
\cite{Metsaev} and so it is exactly solvable \cite{MT2}.

\medskip 

A new soluble sector has been recently proposed by Gomis, Gomis
and Kamimura
\cite{GGK} with the so-called ``non-relativistic limit''
\cite{GO}\footnote{For non-relativistic limit of D-branes in flat space
and AdS space, see
\cite{Brugues:2004an,Gomis:2004pw,Gomis:2005bj,Kamimura:2005rz} and
\cite{Brugues:2006yd,GP}, respectively.}. In this limit the AdS
superstring is reduced to a free theory on the \ads{2} world-sheet with
a static gauge and it is a basically solvable theory. The gauge theory
counterpart has not been clarified yet though some observation has been
given in \cite{GGK}.

\medskip 

The non-relativistic limit is quite similar to the Penrose limit (For
the comparison, see Table\,\ref{vs:tab}).  This similarity leads to a
speculation that the non-relativistic limit can also be described as a
semiclassical limit in the same way as the Penrose limit \cite{GKP}. The
pp-wave string and Penrose limit are recaptured as a semiclassical
approximation of the AdS-superstring around a BPS particle rotating
around the greatest circle in the \s{5} at the speed of light\cite{Frolov:2002av}. 
On the
other hand, in this paper we will show that non-relativistic limit is
nothing but a semiclassical approximation around a ``static''
configuration. Concretely, we will reproduce the non-relativistic
actions of string and D-branes, which have been obtained in
\cite{GGK} and \cite{SY:non-rela} respectively, from the semiclassical
approximation.
Here it should be noted that the non-relativistic string action is
nothing but the semiclassical action expanding around a static \ads{2}
world-sheet previously obtained by Drukker, Gross
and Tseytlin \cite{DGT},
where a 1/2 BPS straight Wilson line \cite{Wilson} is inserted in the
$\mathcal{N}$=4 SYM side. 
What we will show newly is the equivalence 
between 
AdS-brane actions in 
the non-relativistic limit
and 
those in 
the semiclassical limit.
This shows that
a non-relativistic limit is nothing but a semiclassical limit around a
static configuration.

\medskip 

The semiclassical interpretation of the non-relativistic limit leads us
to argue the AdS/CFT dictionary in the non-relativistic limit with the
help of the symmetry argument. Our argument is that a non-relativistic
string state would correspond to small deformation of a 1/2 BPS straight
Wilson line. Unfortunately, it would be difficult to check the
dictionary by directly computing the anomalous dimension with
perturbation theory because no large U(1)$_{\rm R}$ charge is included
unlike the BMN case \cite{BMN} and the BMN scaling technique does not
work at all.

\medskip \medskip

The content of this paper is as follows: 
In section 2 we introduce the Dirac-Born-Infeld (DBI) actions of 
D-branes on the \adss{5}{5}\,. The classical solutions around
which the actions are expanded are also discussed. In section 3 we
reproduce the non-relativistic actions from the semiclassical
approximation. First the bosonic fluctuations are considered for each of
the branes. After that, the fermionic fluctuations are discussed since
these are almost the same for all the branes. 
In section 4 we argue the corresponding gauge-theory side. 
Section 5 is devoted to a summary and discussions. 
Some conventions and notations are collected in Appendix. 

\begin{table}
 \begin{center}
  \begin{tabular}{|c|c|c|} 
\hline
& \quad pp-wave string ~~ & non-relativistic string \\ 
\hline \hline 
limit &  Penrose limit & non-relativistic limit \\ 
gauge & light-cone & static \\ 
8 bosons & 8 massive  & \quad 3 massive $\&$ 5 massless \quad  \\ 
transverse sym.\ & SO(4)$\times$SO(4) & SO(3)$\times$SO(5) \\ 
world sheet  & flat & \ads{2} \\ 
classical sol.\ \ & \quad 
rotating particle ~(1/2 BPS) \quad & static \ads{2} ~(1/2 BPS) \\ 
sym.\ of sol.\ & U(1) & SL(2,R) \\ 
the vacuum op.\ & single trace op.\ Tr$(Z^J)$  &  straight Wilson line  \\ 
\hline
  \end{tabular}
\caption{pp-wave string vs.\ non-relativistic string}\label{vs:tab}
 \end{center}
\end{table}

\section{Setup}

Let us introduce the D-brane actions (including the F-string case) on
the \adss{5}{5} background and discuss static classical solutions around
which the actions are expanded in a semiclassical method.    

\subsection{The actions of strings and D-branes on \adss{5}{5}} 

The D$p$-brane action \cite{D:curved} 
(see also \cite{D:flat})
is composed of the two parts, 
the Dirac-Born-Infeld (DBI) part and the Wess-Zumino (WZ) part as 
follows: 
\begin{eqnarray}
S=S_{\DBI}+S_{\WZ}
=T_p\int_\Sigma(\CL_\DBI+\CL_\WZ)~
\label{S 10}
\end{eqnarray}
where $T_p=({g_s(2\pi)^p\alpha'^{\frac{p+1}{2}}})^{-1}$ is the tension of the D$p$-brane. 

The DBI action is given by\footnote{We suppressed the dilaton and axion
factors here.}  
\begin{eqnarray}
\CL_{\DBI}&=&\sqrt{s\det(g+\CF)}\,d^{p+1}\xi~,
~~~
g_{ij}=\BL_i^A\BL_j^B\eta_{AB}~,~~
\BL^A_i=\partial_iZ^{\hat M}\BL_{\hat M}^A~,~
\end{eqnarray}
where
   $s=-1$ for a Lorentzian
brane while $s=1$ for a Euclidean brane.
$\BL^A$ and $L$ are  supervielbeins on \adss{5}{5} 
written in terms of the superspace coordinates $Z^{\hat M}=(X^M,\theta^\alpha)$
(For the concrete forms see Appendix).  
Then $\CF$ is a modified field strength $\CF=F-\CB$\,. 
$F=dA$ is a gauge field strength and
$\CB$ is (the pullback on the worldvolume
of) the NS-NS two-form
\begin{eqnarray}
\CH=d\CB=i\BL^A\bar L\Gamma_A\sigma L~,~~~
\CB=2i\int^1_0\!\!\!dt\, \hat\BL^A\hat{\bar L}\Gamma_A\sigma\theta
\label{NSNS B}
\end{eqnarray}
where the symbol ``hat'' implies $\hat E=E(t\theta)$\,. 
In the case that we want to consider a single F-string, 
the DBI part should be replaced by the Nambu-Goto (NG) action by turning
off the flux as $S_{\NG}=S_\DBI|_{\CF=0}$ and the tension
should also be replaced by $T=(2\pi\alpha')^{-1}$.

The WZ part 
is characterized by a supersymmetric closed $(p+2)$-form $h_{p+2}$
\begin{eqnarray}
h_{p+2}&=&d\CL_{\WZ}
=\sum_{n=0}\frac{1}{n!}h^{(p+2-2n)}\CF^n~
\label{h_{p+2}}
\end{eqnarray}
with
\begin{eqnarray}
h^{(\ell+2)}&=&\frac{\sqrt{s}}{\ell!}\left[\BL^{A_1}\cdots \BL^{A_{\ell}}
\bar L\Gamma_{A_1\cdots A_\ell}\varrho L 
+\delta_{\ell,3}\frac{i\alpha}{5}(\epsilon_{a_1\cdots a_5}\BL^{a_1}\cdots\BL^{a_5}
-
\epsilon_{a_1'\cdots a_5'}\BL^{a_1'}\cdots\BL^{a_5'}
)
\right]~~~~
\end{eqnarray}
where
$\varrho=(\sigma)^{-\frac{p-3}{2}}i\sigma_2$
with
$\sigma=\sigma_3$ and $-\sigma_1$ for D$p$-brane and F-string, 
respectively.
The $(p+1)$-dimensional form of the WZ term is
\begin{eqnarray}
\CL_{\WZ}
&=& \int_0^1 \!\!d t\,  
[\SC\wedge \e^{\hat\CF}]_{p+1}
+C^{(p+1)}
~,~
\nonumber\\
\SC&=&\bigoplus_n
\frac{2\sqrt{s}}{(2n-1)!}\hat\BL^{A_1}\cdots\hat\BL^{A_{2n-1}}
\hat{\bar L}\Gamma_{A_1\cdots A_{2n-1}}(\sigma)^{n}i\sigma_2\theta~
\end{eqnarray}
where $C^{(p+1)}$ is a bosonic $(p+1)$-form
satisfying $h_{p+2}\big|_{\text{bosonic}}=dC^{(p+1)}$\,.

\subsection{Classical static solutions}

Here let us derive static D-brane solutions. 

First we shall introduce the variation formulae for the supervielbeins
and super spin connections, which can be derived from the MC equations
give in (\ref{MC AdSxS in 10-dim}). The formulae are as follows: 
\begin{eqnarray}
\delta\BL^A&=&
d\delta x^A
-\eta_{BC}\delta x^{AB}\BL^C
+\eta_{BC}\BL^{AB}\delta x^C
-2i\bar L\Gamma^A\delta\theta~,\nonumber\\
\delta L&=&
d\delta\theta
-\frac{\alpha}{2}\delta x^A\widehat\Gamma_Ai\sigma_2 L
+\frac{\alpha}{2}\BL^A\widehat\Gamma_Ai\sigma_2\delta\theta
-\frac{1}{4}\delta x^{AB}\Gamma_{AB}L
+\frac{1}{4}\BL^{AB}\Gamma_{AB}\delta\theta
~,\nonumber\\
\delta\BL^{ab}&=&
d\delta x^{ab}
+2\alpha^2\BL^{a}\delta x^{b}
+2\eta_{cd}\BL^{ac}\delta x^{db}
+2i\alpha\bar L\widehat\Gamma^{ab}i\sigma_2\delta\theta~,
\label{variation L}
\\
\delta\BL^{a'b'}&=&
d\delta x^{a'b'}
-2\alpha^2\BL^{a'}\delta x^{b'}
+2\eta_{c'd'}\BL^{a'c'}\delta x^{d'b'}
+2i\alpha\bar L\widehat\Gamma^{a'b'}i\sigma_2\delta\theta\,, \nonumber 
\end{eqnarray}
where we have introduced the following quantities: 
\begin{eqnarray}
\delta x^A=\delta Z^{\hat M}\BL_{\hat M}^A\,, \quad 
\delta x^{AB}=\delta Z^{\hat M}\BL_{\hat M}^{AB}\,, \quad 
\delta \theta^\alpha=\delta Z^{\hat M}L_{\hat M}^{\alpha}\,. 
\label{delta x definition}
\end{eqnarray}
It also follows from (\ref{NSNS B}) that
\begin{eqnarray}
\delta\CB&=&
i\delta x^A\bar L\Gamma_A\sigma L
+2i\BL^A\bar L\Gamma_A\sigma L
+d\left(
\int^1_0\!\!\!dt\,
[\delta \hat x^A\hat{\bar L}\Gamma_A\sigma\theta
-\hat\BL^A\delta\hat{\bar \theta}\Gamma_A\sigma\theta]
\right)~. \label{delta B}
\end{eqnarray}

\medskip 

Then, by using the variation formulae (\ref{variation L}) and
(\ref{delta B}), 
a variation of the D$p$-brane action is derived as
\begin{eqnarray}
\delta S&=& T_p\int_\Sigma\!\! d^{p+1}\xi\,
\frac{1}{2}\sqrt{s\det(g+\CF)}(g+\CF)^{ji}(\delta g+\delta F-\delta\CB)_{ij}
\nonumber\\&&
+T_p\int_B
\sum_{n=0}\left[
\frac{1}{n!}\delta h^{(p+2-2n)}\CF^n+\frac{1}{(n-1)!}h^{(p+2-2n)}\CF^{n-1}\delta\CF
\right]\,, 
\label{variation}
\end{eqnarray}
where $\partial B=\Sigma$\,. 
For a classical solution this variation should be zero. 

\medskip 

It is the standard procedure to impose some ansatz in finding a
classical solution.  First let us impose that $\theta=0$\,. Then
(\ref{variation}) reduces to
\begin{eqnarray}
\delta S &=& T_p\int_\Sigma \!\! d^{p+1}\xi
\frac{1}{2}\sqrt{s\det(g_B+F)}(g_B+F)^{ji}
\left(
2e^A_i\eta_{AB}
\delta\BL_j^B|_{\theta=0}
+\delta F_{ij}
\right)~
\nonumber\\&&
+T_p\int_B\frac{\sqrt{s}i\alpha}{3!}\Bigg[
\Big(\epsilon_{a_1\cdots a_5}
\delta \BL^{a_1}|_{\theta=0}
e^{a_2}\cdots e^{a_5}
-
\epsilon_{a_1'\cdots a_5'}
\delta \BL^{a_1'}|_{\theta=0}
e^{a_2'}\cdots e^{a_5'}\Big)\frac{1}{(\frac{p-3}{2})!}F^{\frac{p-3}{2}}
\nonumber\\&&~~~~~~
+\frac{1}{5}\Big(\epsilon_{a_1\cdots a_5}e^{a_1}\cdots e^{a_5}
-
\epsilon_{a_1'\cdots a_5'}e^{a_1'}\cdots e^{a_5'}\Big)
\frac{1}{(\frac{p-5}{2})!}F^{\frac{p-5}{2}}d\delta A
\Bigg]\,, \nonumber 
\end{eqnarray}
where $g_{Bij}=e^A_ie^B_j\eta_{AB}$ and
$\delta\BL^A|_{\theta=0}=
d\delta x^A
-\eta_{BC}\delta x^{AB}e^C
+\eta_{BC}\omega^{AB}\delta x^C$\,.

Furthermore we require that $A_i=0$\,. 
Then it further reduces to 
\begin{eqnarray}
\delta S& =&
-T_p\int_\Sigma\!\! d^{p+1}\xi
\sqrt{s\det g_B }g_B^{ji}\left(
\nabla_ie_j^A
+\omega^A_i{}_Be^B_j
\right)
\delta x_A~
\nonumber\\&&
+T_p\int_B\frac{\sqrt{s}i\alpha}{3!}\Bigg[
\delta_{p,3}\Big(
\epsilon_{a_1\cdots a_5}(d\delta x^{a_1}+\eta_{bc}\omega^{a_1b}\delta x^{c})e^{a_2}\cdots e^{a_5}
\nonumber\\&&~~~~~~~~~~~~~
-
\epsilon_{a_1'\cdots a_5'}(d\delta x^{a_1'}+\eta_{b'c'}\omega^{a_1'b'}\delta x^{c'})e^{a_2'}\cdots e^{a_5'}\Big)
\nonumber\\&&
+\frac{1}{5}
\delta_{p,5}(
\epsilon_{a_1\cdots a_5}e^{a_1}e^{a_2}\cdots e^{a_5}
-
\epsilon_{a_1'\cdots a_5'}e^{a_1'}e^{a_2'}\cdots e^{a_5'})d\delta A
\Bigg]\,, \nonumber 
\end{eqnarray}
where we have partially integrated. 
The terms proportional to $\delta_{p,3}$ and $\delta_{p,5}$ 
can be deleted by performing a partial integration and 
using the relation $de^A=-\eta_{BC}\omega^{AB}e^C$\,. 
This is a consequence of $dh_{p+2}=0$\,. 
Then the remaining term (the first line) vanishes when the following
relation is satisfied:  
\begin{eqnarray}
\nabla_ie_j^A+\omega^A_i{}_Be^B_j=0\,. 
\label{eom}
\end{eqnarray} 
For the static configuration $X^{M_i}=\xi^i$ ($i=0,\cdots,p$)
this relation can be derived as the pull-back of the following relation 
to the world-sheet $\Sigma$:  
\begin{eqnarray}
\nabla_Me_N^A +\omega^A_M{}_Be^B_N=0\,, 
\label{bosonic MC}
\end{eqnarray}
which follows from the definition of the spin connection 
$\omega^A_M{}_B=-e^P_B\nabla_Me^A_P$\,. 
Thus the static gauge configuration is a trivial classical solution.  

Besides it, one can find a generalized configuration represented by
\begin{eqnarray}
X^{M_i}
=\left\{
  \begin{array}{ll}
X^{M_i}(\xi^i)       &~~\mathrm{for}~~ i=0,\cdots,p   \\
0       &~~\mathrm{for}~~  i=p+1,\cdots,9  \\ 
  \end{array}
\right.\,. 
\label{classical solution}
\end{eqnarray} 
One can easily check that this solves (\ref{eom}) as follows.
First
we  rewrite (\ref{eom}) as
\begin{eqnarray}
(\nabla_i\nabla_jX^M+\nabla_iX^P\nabla_jX^Q\Gamma_{PQ}^M)e^A_M
=0
\label{eom X}
\end{eqnarray}
where we have used (\ref{bosonic MC}).
Defining $\eta^M_i\equiv \partial_i X^M$
($M\in\{M_i|i=0,\cdots,p\}$)
which is invertible for the classical solution (\ref{classical solution})
one derives by using $g^{kl}\eta_k^M\eta_l^N=G^{MN}$
\begin{eqnarray}
\gamma_{ij}^k\partial_kX^M&=&
(
g^{kl}\eta^P_l\partial_{(i}\eta_{j)}^QG_{PQ}
+g^{kl}\eta^P_i\eta^Q_j\eta^R_lG_{RS}\Gamma^S_{PQ}
)\eta_k^M
\nonumber\\&=&
\partial_i\partial_jX^M
+\partial_iX^P\partial_jX^Q\Gamma^M_{PQ}~
\end{eqnarray}
which solves (\ref{eom X}).
Thus (\ref{classical solution}) is also a static D$p$-brane solution but 
its parametrization of the world-volume coordinates is slightly
generalized. 
Even for the F-string case the same solution can be derived. 
Hereafter we will concentrate on the solution (\ref{classical
solution}). 

\medskip 

The value of the classical action can be evaluated 
by putting the classical solution into the action as follows: 
\begin{eqnarray}
\CL_c&=&\sqrt{s\det g_0}\,d^{p+1}\xi
= \det (e_0)^A_{i} d^{p+1}\xi
= \frac{1}{2}\epsilon_{A_0\cdots A_{p}}e^{A_0}_0\cdots e^{A_p}_0\,,
\label{value}
\end{eqnarray}
where
$g_0=g|_\mathrm{classical}$ and $e_0^A=e^A|_\mathrm{classical}$\,. The
contribution (\ref{value}) 
can be canceled out by adding $C^{(p+1)}$ of the form
\begin{eqnarray}
C^{(p+1)}_0=-\frac{1}{2}\epsilon_{A_0\cdots A_{p}}e^{A_0}_0\cdots
 e^{A_p}_0\,,
\nonumber 
\end{eqnarray}
only if one can find such a closed form
that $h_{p+2}|_\mathrm{classical}=dC^{(p+1)}_0$. 
For the 1/2 BPS D-branes we can
easily show that
\[
h_{p+2}|_\mathrm{classical}=0\,
\] 
and that $dC^{(p+1)}_0=0$,
hence
the classical action (\ref{value}) can be canceled out by the flux
term. This cancellation is actually necessary for the consistent
semiclassical approximation. 
For the semiclassical approximation we need to
take a large tension limit, which surely corresponds to the limit
operation in the non-relativistic limit. Thus, if the cancellation does
not occur, then the classical action diverges. That is why the
cancellation is necessary.

\medskip 

It is helpful to remember a classification of possible 1/2 BPS
configurations of D-branes on the \adss{5}{5} (listed in Table\
\ref{IIB-branes}) \cite{SMT,Sakaguchi:2003py}. 
 The notation of
$(m,n)$-brane implies a brane configuration whose worldvolume extends
along $m$ directions in the \ads{5} and $n$ directions in the \s{5}.
Such a restriction against the directions has its roots in a 1/4 BPS
intersecting condition before taking the
near-horizon limit \cite{SMT}. The classification could be reproduced in
various ways, brane probe analysis \cite{SMT}, $\kappa$-invariance
\cite{Sakaguchi:2003py} and consistent non-relativistic limit
\cite{SY:non-rela} or semiclassical limit as discussed here.

\vspace*{0.5cm}

\begin{table}[htb]
 \begin{center}
  \begin{tabular}{|c|c|c|c|}
    \hline         
1-brane  &3-brane    &5-brane    &7-brane    \\ 
    \hline
  (2,0),~(0,2)   & (3,1),~(1,3)  &(4,2),~(2,4)    &  (5,3),~(3,5)      \\
    \hline
  \end{tabular}
 \end{center}
  \caption{The possible configurations of the 1/2 BPS branes in
\adss{5}{5}} \label{IIB-branes}
\end{table}

\section{Quantum Fluctuations around the Static Solutions} 

From now on we shall consider a semiclassical approximation of the
D-brane action around the classical solutions obtained in the previous
section. As a result we see that the non-relativistic D-brane actions
are reproduced. 

\medskip 

First of all, let us expand the D-brane action in terms of the fermionic
variables $\theta$: 
\begin{eqnarray}
&&
\mathcal{L}_{\rm DBI}
=
\sqrt{s\det(g_B+F)}\left[
1+\frac{1}{2}(g_B+F)^{ji}(g_F-B_F)_{ij}
\right]d^{p+1}\xi +O(\theta^4)\,, \nonumber \\
&&\CL_{\rm WZ}= 
C^{(p+1)} \nonumber  \\ 
&& \qquad\quad +
\sum_{n=1}
\frac{-\sqrt{s}}{(2n-1)!}
e^{A_1}\cdots e^{A_{2n-1}}\bar\theta\Gamma_{A_1\cdots A_{2n-1}}(\sigma)^{n}i\sigma_2
D\theta \frac{F^{p+1-2n}}{(p+1-2n)!}
+O(\theta^4)\,,
\nonumber 
\end{eqnarray}
where $g_B$\,, $g_{F}$ and $B_F$ are 
\begin{eqnarray}
g_B{}_{ij}=e^A_ie^B_j\eta_{AB}\,, \quad 
g_F{}_{ij}=2ie^A_{(i}\bar\theta\Gamma^{B}D_{j)}\theta\eta_{AB}\,, \quad 
B_F=-ie^A\bar\theta\Gamma_A\sigma D\theta\,. \nonumber 
\end{eqnarray}
Then the bosonic and the fermionic parts are given by 
\begin{eqnarray}
S_B&=&  T_p\int\! d^{p+1}d\xi\, \sqrt{s\det(g_B+F)}
+T_p\int_\Sigma C^{(p+1)}\,, \nonumber 
\\
S_F&=&
 T_p\int\! d^{p+1}d\xi\,
\sqrt{s\det(g_B+F)}
\frac{1}{2}(g_B+F)^{ji}(g_F-B_F)_{ij}
\nonumber\\&&
+T_p\int_\Sigma
\sum_{n=1}
\frac{-\sqrt{s}}{(2n-1)!}
e^{A_1}\cdots e^{A_{2n-1}}\bar\theta\Gamma_{A_1\cdots A_{2n-1}}(\sigma)^{n}i\sigma_2
D\theta \frac{F^{p+1-2n}}{(p+1-2n)!}
+O(\theta^4)\,. \nonumber 
\end{eqnarray}

\medskip 

Our purpose is to examine the fluctuations about the classical
solution. Hence let us decompose the variables into the fluctuations
(expressed with ``tilde'') and
the background (\ref{classical solution}) with $\theta=0$ and $A_i=0$ as
follows: 
\begin{eqnarray}
X^{\Mu_i}=X^{\Mu_i}_0(\xi^i)+\tilde X^{\Mu_i}\,, \quad 
X^{\Mb_i}=0+\tilde X^{\Mb_i}\,, \quad 
\theta= 0+\tilde\theta\,, \quad 
A_i= 0+\tilde A_i\,, \nonumber 
\end{eqnarray}
where $\Mu=\{\,M_i~|~i=0,\cdots,p\,\}$ and $\Mb=\{\,M_i~|~i=p+1,\cdots,9\,\}$\,. 

\medskip 

Here we should remark on the treatment of the longitudinal fluctuations.
In order to maintain the static gauge form (\ref{classical solution}),
we have to restrict the longitudinal fluctuation $\tilde X^{\Mu_i}$ to
be $\tilde X^{\Mu_i}=\tilde X^{\Mu_i}(\xi^i)$\,.  Then, $\tilde
X^{\Mu_i}(\xi^i)$ however can be absorbed into $X^{\Mu_i}_0(\xi^i)$, and
hence $\tilde X^{\Mu_i}$ does not appear in the semiclassical action
after all. For F-string case it has already been discussed in
\cite{DGT}. The vanishing of the longitudinal modes for the F-string is
plausible also from the equivalence between the Nambu-Goto action and
Polyakov action. In the Polyakov action the longitudinal modes should be
canceled by conformal ghosts while the ghosts are not included in the
Nambu-Goto action.  But in the case of the Nambu-Goto this cancellation
comes from the requirement that the static gauge should be maintained.
Thus, anyway, it is sufficient
to consider the transverse fluctuations.

\medskip

We should comment on the scaling of brane tension.  The common radius 
of \ads{5} and \s{5}, $R$\,, can be factored out to be an overall factor of
the action.  This may replace $T_p$ by
\begin{eqnarray}
T_pR^{p+1}=t_p\lambda^{\frac{p+1}{4}}~,~~~
t_p=(g_s(2\pi)^\frac{p-1}{2})^{-1}~,~~~
\sqrt{\lambda}\equiv \frac{R^2}{2\pi\alpha'}~.
\end{eqnarray}
We make $\lambda^{\frac{p+1}{4}}$ be absorbed 
into the fluctuations $\tilde Z$ as $\tilde Z\to\lambda^{-\frac{p+1}{8}}\tilde Z$,
thus considering quadratic fluctuations
is equivalent to considering the large $\lambda$ limit.
The cancellation of the classical contributions is necessary
for the consistent large $\lambda$ limit. \footnote{
Now one sees the direct correspondence between
the semiclassical limit and
the non-relativistic limit arranged in \cite{SY:non-rela}
(see equations (5.1) and (5.2) there).
}

\medskip 

We will see that the semiclassical AdS-brane actions completely agree
with the non-relativistic AdS-brane actions derived in \cite{SY:non-rela}.
The bosonic and the fermionic fluctuations will be separately discussed
below. First the bosonic ones are evaluated for each of the D-brane
configurations. Then the fermionic ones are done in a unified way.

\subsection{Bosonic fluctuations}

From now on let us consider the bosonic fluctuations.  Hereafter the
world-sheet Wick rotation is assumed when we take a Euclidean brane with
$s=1$\,.  The metric of \adss{5}{5} we take is given in (\ref{metric:
AdS5}) and (\ref{metric: S5}).

\subsubsection*{D-string (F-string)}

The bosonic part of the action is given by 
\begin{eqnarray}
\CL_B&=&\sqrt{s\det(g_B+F)}\,d^2\xi\,.
\label{L D1 bosonic}
\end{eqnarray}
Let us consider the \ads{2}-brane (i.e., (2,0)-brane) and 
expand (\ref{L D1 bosonic}) around 
the classical solution
\begin{eqnarray}
t=t(\xi^0)~,~~~\rho=\rho(\xi^1)
\,. \nonumber 
\end{eqnarray}
Here $t$ and $\rho$ depend only on $\xi^0$ and $\xi^1$\,, respectively. 
The induced metric is 
\begin{eqnarray}
g_0{}_{ij}&=&
-\cosh^2\rho\partial_it\partial_jt
+\partial_i\rho\partial_j\rho~
=\diag(-\cosh^2\rho(\partial_0 t)^2,
(\partial_1\rho)^2)\,, \nonumber 
\end{eqnarray}
and the world-sheet geometry is \ads{2} rather than flat. It is an easy
task to derive 
\begin{eqnarray}
\sqrt{s\det g_0}=\cosh\rho\partial_0t\partial_1\rho\,, \qquad 
g_0^{ij}=\diag(-\frac{1}{\cosh^2\rho(\partial_0 t)^2},
\frac{1}{(\partial_1\rho)^2})\,. \nonumber 
\end{eqnarray}
Since the induced metric is expanded as follows: 
\begin{eqnarray}
g_B{}_{ij}&=&g_0{}_{ij}+g_2{}_{ij}+\cdots\,, \qquad 
g_2{}_{ij}=
\sum_{p=1}^3\sinh^2\rho\partial_i\tilde\phi_p\partial_j\tilde\phi_p
+\partial_i\tilde\gamma\partial_j\tilde\gamma
+\sum_{q=1}^4\partial_i\tilde\varphi_q\partial_j\tilde\varphi_q
\,, \nonumber \\
F&=&d\tilde A\equiv F_1\,, \nonumber 
\end{eqnarray}
the quadratic part of $\CL_B$ is given by 
\begin{eqnarray}
\CL_{2B}&=&\frac{1}{2}\sqrt{s\det g_0}\,
\left[g_0^{ij}g_2{}_{ij}+\frac{1}{2}F_1{}_{ij}F_1^{ij}\right]\,, \nonumber \\ 
&=& \frac{1}{2}\sqrt{s\det g_0}\,\left[
g_0^{ij}(\sinh^2\rho\partial_i\tilde\phi_p\partial_j\tilde\phi_p
+\partial_i\tilde\gamma\partial_j\tilde\gamma
+\partial_i\tilde\varphi_q\partial_j\tilde\varphi_q)
+\frac{1}{2}F_1{}_{ij}F_1^{ij}
\right]\,. \label{actionD}
\end{eqnarray}
By rescaling $\tilde\phi_p$ as $\bar\phi_p=\sinh\rho\tilde\phi_p$\,,  the
first term in (\ref{actionD}) can be rewritten as 
\begin{eqnarray}
&&\frac{1}{2}
\sqrt{s\det g_0}\,
g_0^{ij}\sinh^2\rho\partial_i\tilde\phi_p\partial_j\tilde\phi_p
=\frac{1}{2}
\sqrt{s\det g_0}\,\left[
g_0^{ij}\partial\bar\phi_p\partial\bar\phi_p
+2\bar\phi^2_p \right]\,, \nonumber  
\end{eqnarray}
where a partial integration has been performed. 
Thus one can obtain that 
\begin{eqnarray}
S_{2B}
&=&t_1\int \! d^2\xi
\sqrt{s\det g_0}\,\left[\frac{1}{2}
g_0^{ij}(\partial_i\bar\phi_p\partial_j\bar\phi_p
+\partial_i\tilde\gamma\partial_j\tilde\gamma
+\partial_i\tilde\varphi_q\partial_j\tilde\varphi_q)
+\bar\phi^2_p
+\frac{1}{4}F_1{}_{ij}F_1^{ij}
\right]\,
\label{fluctuation action:bosonic}
\end{eqnarray}
where $R$ is absorbed into fluctuations.
This is just the non-relativistic AdS D-string action derived in 
\cite{SY:non-rela}.
This reduces to the non-relativistic AdS F-string action derived in 
\cite{DGT, GGK} by setting $\tilde{A}$ to be zero.

\subsubsection*{D3-brane}

The bosonic part of the D3-brane action is\footnote{The supersymmetric 
actions of D3-branes on the \adss{5}{5} and the pp-wave are
constructed in \cite{D3:AdS} and \cite{D3:pp}, respectively. }
\begin{eqnarray}
\CL_B&=&
\sqrt{s\det(g_B+F)}\,d^4\xi
+C^{(4)}~. \nonumber 
\end{eqnarray}
Let us consider the \adss{3}{1}-brane (i.e., (3,1)-brane) first. 
The classical solution is given by 
\begin{eqnarray}
(t,\rho,\phi_1;\gamma)=(t(\xi^0),\rho(\xi^1),\phi_1(\xi^2);\gamma(\xi^3))~.
 \nonumber 
\end{eqnarray}
and the induced metric can be expanded around it as
\begin{eqnarray}
g_B&=&g_0+g_2+\cdots\,,  \nonumber \\
g_0{}_{ij}&=&\diag(-\cosh^2\rho(\partial_0t)^2,
(\partial_1\rho)^2,
\sinh^2\rho(\partial_2\phi_1)^2,
(\partial_3\gamma)^2
)\,, \nonumber \\
g_2{}_{ij}&=&\sinh^2\rho\cos^2\phi_1
\sum_{p=2,3}\partial_i\tilde\phi_p\partial_j\tilde\phi_p
+\cos^2\gamma
\sum_{q=1}^4
\partial_i\tilde\varphi_q\partial_j\tilde\varphi_q\,. \nonumber 
\end{eqnarray}
Defining the new angle variables: 
\begin{eqnarray}
\bar\phi_p\equiv\sinh\rho\cos\phi_1\tilde\phi_p\,, \qquad 
\bar\varphi_q\equiv \cos\gamma\tilde\varphi_q\,, \nonumber 
\end{eqnarray}
the bosonic quadratic action of the DBI part can be rewritten as 
\begin{eqnarray}
&& \frac{1}{2}\sqrt{s\det(g_B+F)}\,  g_0^{ij}g_2{}_{ij} \nonumber \\ 
&& \hspace*{1cm} 
=\sqrt{s\det g_0}\,\left[\,
\frac{1}{2}g_0^{ij}(\partial_i\bar\phi_p\partial_j\bar\phi_p
+\partial_i\bar\varphi_q\partial_j\bar\varphi_q
)+\frac{1}{2}(3\bar\phi^2_p-\bar\varphi^2_q)
+\frac{1}{4}F_1{}_{ij}F_1^{ij}\,
\right]\,, \label{res1}
\end{eqnarray}
where we have performed partial integrations. Then it is turn to
consider the WZ part.  The $C^{(4)}$ is now expressed as 
\begin{eqnarray}
C^{(4)}&=&4\sqrt{s}i(
-\cosh\rho\sinh^3\rho\cos^2\phi_1
dtd\rho d\phi_1 \phi_2 d\phi_3
+\cos^4\gamma d\gamma\varphi_1d\varphi_2\cdots d\varphi_4)\,, \nonumber 
\end{eqnarray}
up to an exact term, and it can be expanded in terms of the fluctuations
as 
\begin{eqnarray}
C^{(4)}&=&C^{(4)}_2+\cdots\,, \nonumber 
\end{eqnarray}
and the integration of the quadratic part is given by 
\begin{eqnarray}
\int C^{(4)}_2&=&-
4\sqrt{s}i\int
\cosh\rho\sinh^3\rho\cos^2\phi_1
dtd\rho d\phi_1 \tilde\phi_2 d\tilde\phi_3
\nonumber \\ &=& 
-4\sqrt{s}i\int\! d^4\xi\,
\cosh\rho\sinh\rho
\partial_0t\partial_1\rho \partial_2\phi_1 \bar\phi_2 \partial_3\bar\phi_3
\nonumber\\&=&
-4\sqrt{s}i\int_\Sigma\!
\vol_{\Sigma_3}\, \bar\phi_2 d\bar\phi_3\,,  \label{res2}
\end{eqnarray}
where $\Sigma_3$ is the worldvolume extending in the \ads{5}.  Combining
(\ref{res1}) and (\ref{res2})\,, one can obtain the following quadratic
action
\begin{eqnarray}
S_{2B}&=&
t_3 \int\! d^4\xi\,
\sqrt{s\det g_0}\,\left[\,
\frac{1}{2}g_0^{ij}(\partial_i\bar\phi_p\partial_j\bar\phi_p
+\partial_i\bar\varphi_q\partial_j\bar\varphi_q
)+\frac{1}{2}(3\bar\phi^2_p-\bar\varphi^2_q)
+\frac{1}{4}F_1{}_{ij}F_1^{ij}\,\right]
\nonumber\\&&
-4\sqrt{s}i t_3\int_\Sigma
\vol_{\Sigma_3} \bar\phi_2 d\bar\phi_3~.
\end{eqnarray}
This is nothing but the non-relativistic AdS D3-brane action derived in 
\cite{SY:non-rela}.

\medskip 

On the other hand, we may consider (1,3)-brane. The shape is actually
$R\times$S$^3$ and so
this is related to the
giant graviton \cite{giant} rather than AdS-branes.  The static solution
is given by
\begin{eqnarray}
(t;\gamma,\varphi_1,\varphi_2)
=(t(\xi^0);\gamma(\xi^1),\varphi_1(\xi^2),\varphi_2(\xi^3)) \nonumber 
\end{eqnarray}
as it corresponds to a giant graviton. By carrying out the same
analysis, we can obtain the following quadratic action: 
\begin{eqnarray}
S_{2B}&=& 
t_3 \int\!\! d^4\xi\, 
\sqrt{s\det g_0}\,\left[\,\frac{1}{2}g_0^{ij}(
\partial_i\tilde\rho\partial_j\tilde\rho
+
\partial_i\bar\phi_p\partial_j\bar\phi_p
+
\partial_i\bar\varphi_q\partial_j\bar\varphi_q
)\right. \nonumber \\ 
&& \qquad 
\left. +\frac{1}{2}(\tilde\rho^2+\bar\phi^2_p-3\bar\varphi_q^2)
+\frac{1}{4}F_1{}_{ij}F_1^{ij}\,
\right] 
+4\sqrt{s}it_3  \int_\Sigma\!
\vol_{\Sigma_3'}\, \bar\varphi_3 d\bar\varphi_4~\,. 
\end{eqnarray}
where 
$p=1,2,3$ and $q=3,4$. 
$\Sigma'_3$ is the worldvolume expanding in the \s{5}.

\subsubsection*{D5-brane}

The bosonic part of the D5-brane action is
\begin{eqnarray}
\CL_B&=&
\sqrt{s\det(g_B+F)}\,d^6\xi
+C^{(6)}\,.  \nonumber 
\end{eqnarray}
Let us consider the \adss{4}{2}-brane (i.e., (4,2)-brane) first. 
The classical solution is 
\begin{eqnarray}
(t,\rho,\phi_1,\phi_2;\gamma,\varphi_1)=(t(\xi^0),\rho(\xi^1),\phi_1(\xi^2),
\phi_2(\xi^3); \gamma(\xi^4),\varphi_1(\xi^5))\,. \nonumber
\end{eqnarray}
The induced metric is expanded as
\begin{eqnarray}
g_B&=&g_0+g_2+\cdots\,, \nonumber 
\end{eqnarray}
and the zeroth and the second order parts are given by, respectively, 
\begin{eqnarray}
g_0{}_{ij}&=&\diag(-\cosh^2\rho(\partial_0t)^2,
(\partial_1\rho)^2,
\sinh^2\rho(\partial_2\phi_1)^2,
\sinh^2\rho\cos^2\phi_1(\partial_3\phi_2)^2,
\nonumber\\&&~~~~~~~
(\partial_4\gamma)^2,
\cos^2\gamma(\partial_5\varphi_1)^2
)\,, \nonumber \\
g_2{}_{ij}&=&
\sinh^2\rho\cos^2\phi_1\cos^2\phi_2\partial_i\tilde\phi_3\partial_j\tilde\phi_3
+\sum_{q=2}^4\cos^2\gamma\cos^2\varphi_1
\partial_i\tilde\varphi_q\partial_j\tilde\varphi_q\,. \nonumber 
\end{eqnarray}
Defining the new variables
\begin{eqnarray}
\bar\phi_3 \equiv\sinh\rho\cos\phi_1\cos\phi_2\tilde\phi_3\,, \quad 
\bar\varphi_q\equiv \cos\gamma\cos\varphi_1\tilde\varphi_q\,, \nonumber 
\end{eqnarray}
one can rewrite the quadratic part of the bosonic DBI action as 
\begin{eqnarray}
&&\frac{1}{2}\sqrt{s\det(g_B+F)}\, g_0^{ij}g_2{}_{ij}
\nonumber\\&&
=\sqrt{s\det g_0}\,\left[
\frac{1}{2}g_0^{ij}(\partial_i\bar\phi_3\partial_j\bar\phi_3
+\partial_i\bar\varphi_q\partial_j\bar\varphi_q
)+2\bar\phi^2_3-\bar\varphi^2_q
+\frac{1}{4}F_1{}_{ij}F_1^{ij} 
\right]\,, \label{D5-res1}
\end{eqnarray}
where we have performed partial integrations. 
Then let us consider the WZ part. Since in the present metric we have  
\begin{eqnarray}
dC^{(6)}=h_7|_\mathrm{bosonic}=\frac{\sqrt{s}i}{3!5}(
\epsilon_{a_1\cdots a_5}e^{a_1}\cdots e^{a_5}
-
\epsilon_{a_1'\cdots a_5'}e^{a_1'}\cdots e^{a_5'}
)F\,,  \nonumber 
\end{eqnarray}
the $C^{(6)}$ is expanded as
\begin{eqnarray}
C^{(6)}&=&C^{(6)}_2+\cdots\,. \nonumber 
\end{eqnarray}
The integration of $C^{(6)}_2$ is given by 
\begin{eqnarray}
\int C^{(6)}_2&=&
4\sqrt{s}i\int\!
\cosh\rho\sinh^3\rho\cos^2\phi_1\cos\phi_2
dtd\rho d\phi_1 d\phi_2 \tilde\phi_3~F_1
\nonumber\\&=& 
4\sqrt{s}i\int_\Sigma\!
\vol_{\Sigma_4}\, \bar\phi_3 F_1\,, \label{D5-res2}
\end{eqnarray}
where $\vol_{\Sigma_4}= \cosh\rho\sinh^2\rho\cos\phi_1 dtd\rho d\phi_1
d\phi_2$\,.  Combining (\ref{D5-res1}) and (\ref{D5-res2}), one can find
the quadratic action
\begin{eqnarray}
S_{2B}&=& 
t_5 \int\! d^6\xi\,
\sqrt{s\det g_0}\,\left[
\frac{1}{2}g_0^{ij}(\partial_i\bar\phi_3\partial_j\bar\phi_3
+\partial_i\bar\varphi_q\partial_j\bar\varphi_q
)+2\bar\phi^2_3-\bar\varphi^2_q
+\frac{1}{4}F_1{}_{ij}F_1^{ij}
\right] \nonumber \\ 
&& +4\sqrt{s}i t_5 \int_\Sigma\!\vol_{\Sigma_4}\, \bar\phi_3 F_1\,. 
\end{eqnarray}
This is nothing but the non-relativistic AdS D5-brane action derived in 
\cite{SY:non-rela}.

\medskip 

It is also interesting to consider the \adss{2}{4}-brane case\footnote{
This shape may be related to a giant Wilson loop \cite{gWilson}.}. The
classical solution is given by
\begin{eqnarray}
(t,\rho;\gamma,\varphi_1,\varphi_2,\varphi_3)
=(t(\xi^0),\rho(\xi^0);\gamma(\xi^0),\varphi_1(\xi^0),\varphi_2(\xi^0),\varphi_3(\xi^0))\,.  \nonumber 
\end{eqnarray}
By following the same line, we can obtain the following quadratic 
action
\begin{eqnarray}
S_{2B}&=& t_5 \int\! d^6\xi\,
\sqrt{s\det g_0}\, \left[
\frac{1}{2}g_0^{ij}(
\partial_i\bar\phi_p\partial_j\bar\phi_p
+\partial_i\bar\varphi_4\partial_j\bar\varphi_4
)+\bar\phi^2_p-2\bar\varphi_4^2
+\frac{1}{4}F_1{}_{ij}F_1^{ij} \right] \nonumber \\ 
&& -4\sqrt{s}i t_5 \int_\Sigma\!
\vol_{\Sigma_4'}\, \bar\varphi_4 F_1\,
\end{eqnarray}
where $p=1,2,3$.
This also agrees with the result of \cite{SY:non-rela}.

\subsubsection*{D7-brane}

The bosonic part of the D5-brane action is
\begin{eqnarray}
\CL_B&=& \sqrt{s\det(g_B+F)}\,d^8\xi +C^{(8)}\,.  \nonumber 
\end{eqnarray}
Let us consider the \adss{5}{3}-brane (i.e., (5,3)-brane)
classical solution
\begin{eqnarray}
(t,\rho,\phi_1,\phi_2,\phi_3;\gamma,\varphi_1,\varphi_2)
=(t(\xi^0),\rho(\xi^1),
\phi_1(\xi^2),\phi_2(\xi^3),\phi_3(\xi^4);
\gamma(\xi^5),\varphi_1(\xi^6),\varphi_2(\xi^7))\,. \nonumber 
\end{eqnarray}
The induced metric is expanded as
\begin{eqnarray}
g_B&=&g_0+g_2+\cdots\,, \nonumber  
\end{eqnarray}
and the zeroth and the second order parts are given by, respectively, 
\begin{eqnarray}
g_0{}_{ij}&=&\diag(-\cosh^2\rho(\partial_0t)^2,
(\partial_1\rho)^2,
\sinh^2\rho(\partial_2\phi_1)^2,
\sinh^2\rho\cos^2\phi_1(\partial_3\phi_2)^2,
\nonumber \\ 
&&~\sinh^2\rho\cos^2\phi_1\cos^2\phi_2(\partial_4\phi_3)^2,
(\partial_5\gamma)^2,
\cos^2\gamma(\partial_6\varphi_1)^2,
\cos^2\gamma\cos^2\varphi_1(\partial_7\varphi_2)^2
)\,, \nonumber  \\
g_2{}_{ij}&=&
\sum_{q=3,4}
\cos^2\gamma\cos^2\varphi_1\cos^2\varphi_2
\partial_i\tilde\varphi_q\partial_j\tilde\varphi_q\,. \nonumber 
\end{eqnarray}
Defining the new variable, 
$\bar\varphi_q\equiv
\cos\gamma\cos\varphi_1\cos\varphi_2\tilde\varphi_q$\,,  
one can rewrite the quadratic part of the bosonic DBI part as 
\begin{eqnarray}
&&\frac{1}{2}\sqrt{s\det(g_B+F)}\,  g_0^{ij}g_2{}_{ij}
=\sqrt{s\det g_0}\,\left[
\frac{1}{2}g_0^{ij}
\partial_i\bar\varphi_q\partial_j\bar\varphi_q
-\frac{3}{2}\bar\varphi^2_q
+\frac{1}{4}F_1{}_{ij}F_1^{ij}
\right]\,, \nonumber 
\end{eqnarray}
where we have performed partial integrations. 

Then let us consider the WZ part. 
We can start from the following expression,  
\begin{eqnarray}
dC^{(8)}=h_9|_\mathrm{bosonic}=\frac{\sqrt{s}i}{2\cdot3!\cdot 5}
\Bigl(
\epsilon_{a_1\cdots a_5}e^{a_1}\cdots e^{a_5}
-
\epsilon_{a_1'\cdots a_5'}e^{a_1'}\cdots e^{a_5'}
\Bigr)F^2\,, \nonumber 
\end{eqnarray}
and $C^{(8)}$ can be expanded as
\begin{eqnarray}
C^{(8)}=C^{(8)}_2+\cdots\,. \nonumber 
\end{eqnarray}
The integration of the quadratic part is given by 
\begin{eqnarray}
&& \int C^{(8)}_2=
-2\sqrt{s}i\int_\Sigma
\vol_{\Sigma_5} \tilde A F_1\,, \nonumber \\ 
&& \qquad \vol_{\Sigma_5}=
\cosh\rho\sinh^3\rho\cos^2\phi_1\cos\phi_2
dtd\rho d\phi_1 d\phi_2d\phi_3\,.
\nonumber 
\end{eqnarray}
Combining the results we find the fluctuation action
\begin{eqnarray}
S_{2B}&=& 
t_7 \int\! d^8\xi\, 
\sqrt{s\det g_0}\, \left[
\frac{1}{2}g_0^{ij}\partial_i\bar\varphi_q\partial_j\bar\varphi_q
-\frac{3}{2}\bar\varphi^2_q
+\frac{1}{4}F_1{}_{ij}F_1^{ij} \right] \nonumber \\ 
&& -2\sqrt{s}i t_7 \int_\Sigma\! \vol_{\Sigma_5}\, \tilde A F_1\,.
\end{eqnarray}
This is nothing but the non-relativistic AdS D7-brane action derived in
\cite{SY:non-rela}.

\medskip \medskip 

Next let us consider the \adss{3}{5}-brane case, which is slightly
different from the \adss{5}{3}-brane case. For \adss{3}{5}-brane case, the
classical solution is given by 
\begin{eqnarray}
(t,\rho,\phi_1;\gamma,\varphi_1,\varphi_2,\varphi_3,\varphi_4)
=(t(\xi^0),\rho(\xi^1),\phi_1(\xi^2);\gamma(\xi^3),\varphi_1(\xi^4),\varphi_2(\xi^5),
\varphi_3(\xi^6),\varphi_4(\xi^7))\,, \nonumber 
\end{eqnarray}
and we can derive the following action: 
\begin{eqnarray}
S_{2B}&=& 
t_7 \int\! d^8\xi\, 
\sqrt{s\det g_0}\, \left[
\frac{1}{2}g_0^{ij}\partial_i\bar\phi_p\partial_j\bar\phi_p
+\frac{3}{2}\bar\phi_p^2
+\frac{1}{4}F_1{}_{ij}F_1^{ij} \right] \nonumber \\ 
&& +2\sqrt{s}i t_7  \int_\Sigma \! 
\vol_{\Sigma_5'}\, \tilde A F_1~
\end{eqnarray}
where $p=2,3$.
This is also nothing but the non-relativistic AdS D7-brane action
derived in \cite{SY:non-rela}. Here it should be noted that no 
tachyonic mode is contained in the mass spectrum. 
This fact would basically be based on the topological reason that 
the S$^5$ part of the brane is wrapped around the S$^5$ part of the
\adss{5}{5} background. It should stabilize the solution 
even if the supersymmetries are broken due to some effect.

\subsection{Fermionic fluctuations}

Next we shall consider the fermionic fluctuations. As mentioned before,
the fermionic part can be discussed in a unified way in comparison to
the bosonic part, i.e., it does not almost depend on the dimensionality
of D$p$-branes. Hence we will discuss the fermionic part of all D-branes
at a time. 

\medskip  

Expanding about the classical solution, we can derive the quadratic
action for the fermionic variables: 
\begin{eqnarray}
S_{2F} &=& T_p
\int \! d^{p+1}\xi\, 
\sqrt{s\det g_0}\,g_0^{ij}i\bar{\tilde \theta}\gamma_i(D_j\tilde\theta)_0 \nonumber \\ 
&& \qquad - T_p\int_\Sigma \! 
\frac{\sqrt{s}}{p!}e^{A_1}_0\cdots e^{A_p}_0 \bar{\tilde\theta}
\Gamma_{A_1\cdots A_p}(\sigma)^{\frac{p+1}{2}}i\sigma_2(D\tilde\theta)_0\,, 
\nonumber 
\end{eqnarray}
where the covariant derivative is defined as 
\begin{eqnarray}
(D\tilde\theta)_0
=d\tilde\theta+\frac{1}{2}e^A_0\widehat\Gamma_A i\sigma_2\tilde\theta
+\frac{1}{4}\omega^{AB}_0\Gamma_{AB}\tilde\theta\,, \qquad 
\gamma_i=(e_0)^{A}_i\Gamma_A\, \nonumber 
\end{eqnarray}
where subscript $0$ means classical value.
The fermionic action $S_{2F}$ is invariant under the
following fermionic symmetry
\begin{eqnarray}
\delta_\kappa\tilde\theta&=&(1+\Gamma_0)\kappa~,~~~\nonumber\\
\Gamma_0&=&
\frac{s\sqrt{-s}}{\sqrt{s\det g_0}}
(\sigma)^{n-\frac{p-3}{2}}i\sigma_2
\frac{1}{(p+1)!}\epsilon^{i_1\cdots i_{p+1}}
(e_0)^{A_0}_{i_1}\cdots(e_0)^{A_p}_{i_{p+1}}
\Gamma_{A_0\cdots A_{p}}\,, 
\nonumber\\
&=&\sqrt{-s}\Gamma^{A_0\cdots A_p}\rho
\equiv M~,~~~\rho=\left\{
  \begin{array}{ll}
      \sigma_1 &\mathrm{for}~p=1\mod 4    \\
      i\sigma_2 &\mathrm{for}~ p=3\mod 4   \\
  \end{array}
\right.\,.  \label{fer}
\end{eqnarray}
This symmetry is inherited from
the $\kappa$-symmetry of D$p$-brane action
\begin{eqnarray}
\delta_\kappa\theta&=&(1+\Gamma)\kappa~,~~~\nonumber\\
\Gamma&=&\frac{s\sqrt{-s}}{\sqrt{s\det (g+\CF)}}
\sum_{n=0}\frac{1}{2^nn!}
\Gamma^{j_1k_1\cdots j_nk_n}\CF_{j_1k_1}\cdots \CF_{j_nk_n}
\nonumber\\&&
\times(-1)^n
(\sigma)^{n-\frac{p-3}{2}}i\sigma_2
\frac{1}{(p+1)!}\epsilon^{i_1\cdots i_{p+1}}
\Gamma_{i_1\cdots i_{p+1}}\,, \nonumber 
\end{eqnarray}
where $\Gamma_i=\BL_i^A\Gamma_A$\,. 
Fixing the fermionic symmetry (\ref{fer}) with the condition 
\begin{eqnarray}
\tilde\theta_+=0\,, \quad \tilde\theta_\pm=P_\pm\tilde\theta_\pm\,,
 \quad P_\pm=\frac{1}{2}(1\pm M)\,, \nonumber 
\end{eqnarray}
we obtain the gauge fixed action:
\begin{eqnarray}
S_{2F}&=& t_p\int\! d^{p+1}\xi\, \sqrt{s\det g_0}~
2i g_0^{ij}\bar{\vartheta}\gamma_i(D_j\vartheta)_0\,,
\nonumber 
\end{eqnarray}
where we have absorbed $R$ by rescaling $\tilde \theta$\,
and denoted $\vartheta \equiv \tilde \theta_-$.
  This is the
fermionic part of non-relativistic AdS D$p$-brane action derived in
\cite{SY:non-rela}.

\medskip 

Finally we comment on the mass term contained in $(D\vartheta)_0$\,. 
The vielbein and spin connection on the $(p+1)$-dimensional world-volume
for our classical solutions is given by
\begin{eqnarray}
\hat e^\alpha_i=\partial_iX_0^M(e_0)^\alpha_M
\,, \quad 
\hat\omega^{\alpha\beta}_i=\partial_iX_0^M(\omega_0)_M^{\alpha\beta}\,, 
\nonumber 
\end{eqnarray}
where $\alpha$ is the tangential-vector index.
By using the $(p+1)$-dimensional spinorial derivative defined by
\begin{eqnarray}
\nabla_i=\partial_i+\frac{1}{4}\hat\omega^{\alpha\beta}_i
\Gamma_{\alpha\beta}\,, \nonumber 
\end{eqnarray}
one can find that 
\begin{eqnarray}
(D_i\vartheta)_0
=\nabla_i\vartheta
+\frac{1}{2}(e_0)^A_i\widehat\Gamma_A i\sigma_2\vartheta\,.  \nonumber 
\end{eqnarray}
So for $(m,n)$-branes
\begin{eqnarray}
g_0^{ij}\bar{\vartheta}\gamma_i
\frac{1}{2}(e_0)^A_j\widehat\Gamma_A i\sigma_2\vartheta
=
-\frac{m+n}{2}g_0^{ij}\bar{\vartheta}
\CI i\sigma_2\vartheta\,, \nonumber 
\end{eqnarray}
where we have used $\CI\theta=-\CJ\theta$
which follows from $\Gamma_{01\cdots 9}\theta=\theta$.
Thus
\begin{eqnarray}
S_{2F}&=& t_p \int\! d^{p+1}\xi\, \sqrt{s\det g_0}~2i\left[
 g_0^{ij}\bar\vartheta\gamma_i\nabla_j\vartheta
-\frac{m+n}{2}
\bar{\vartheta}
\CI i\sigma_2\vartheta
\right]\,.
\label{fluctuation action:fermionic}
\end{eqnarray}
This is the action for 8 massive fermions propagating on \adss{m}{n}
for the $(m,n)$-branes. 

\medskip 

Thus we have shown that the non-relativistic actions constructed in
\cite{SY:non-rela} can be reproduced from the semiclassical limit around
static 1/2 BPS D-brane configurations.

\section{The Gauge Theory Side}

In the previous section we have shown that the non-relativistic limit 
is nothing but a semiclassical approximation around a static configuration. 
In this section, we argue the corresponding composite operators in the
gauge theory side, according to the semiclassical interpretation for the
non-relativistic limit.

\subsection{BMN Dictionary}

Before considering the non-relativistic limit, it is helpful to remember
the pp-wave case \cite{BMN,GKP}. The pp-wave string can be seen as a
semiclassical approximation around a BPS particle solution rotating at
the velocity of light. The solution has a U(1) charge $J$ associated
with the rotation. Then in the gauge theory side it corresponds to a
single trace operator
\begin{eqnarray}
{\rm Tr}(Z^J)\,, \qquad Z \equiv \phi_5 + i\phi_6\,.  \label{vacuum}
\end{eqnarray}
The solution is invariant under SO(4)$\times$SO(4) and the
fluctuations are represented by insertions of impurities:
\[
 D_{\mu}Z \quad (\mu=0,1,2,3)\,, \qquad \phi_I \quad (I=1,2,3,4)\,,  
\]
into the vacuum operator (\ref{vacuum}). Here it should be noted that
$D_{\mu}Z$ has bare scaling dimension 2 but $Z$ carries the U(1) charge 
1, and hence $\Delta -J =1$\,.

\subsection{Dictionary in Non-relativistic Limit}

Let us consider the non-relativistic string theory. The world-sheet
geometry is \ads{2} and it ends on the AdS boundary. Hence a straight
Wilson line\footnote{Here we discuss a single
\ads{2}-brane and the representation of the corresponding Wilson line is
represented by a single box in terms of Young tableau.} 
\begin{eqnarray}
W = {\rm Tr} ~{\rm P}\exp\left[\int\! dt\,\left(i A_0 + \phi \right)\right]\,, 
\qquad \phi = \sum_{i=1}^6n^i\phi_i\,. \nonumber 
\end{eqnarray} 
is contained in the gauge theory side.  This is 1/2 BPS and invariant
under SO(3)$\times$SO(5). Hereafter we will fix the six-dimensional vector
$n^i$ as $n^i=(0,\ldots,0,1)$\,.

In analogy with the semiclassical interpretation for the Penrose limit
\cite{GKP}, it is plausible to identify the Wilson line with the vacuum
operator. From the mass spectrum of the non-relativistic string, its
transverse symmetry is SO(3)$\times$SO(5) and it should be related to
the fluctuations. Then we argue that the fluctuations would be
represented by the insertions of the following impurities: 
\begin{eqnarray}
D_{a}\phi_6 + iF_{a0} \quad (a=1,2,3)\,, \qquad \phi^{a'} 
\quad (a'=1,\ldots,5)\,. \label{dic}
\end{eqnarray}
For the AdS directions, the field strength is included in comparison to
the BMN case. Here we should note the difference of the scaling
dimensions 1 between \ads{5} and \s{5} directions. 
It is closly related to the difference between the masses of
non-relativistic string as we will see later. 

\medskip 

We will see some supports for the dictionary (\ref{dic}) below.

\subsection{Wilson Loop Expansion}

The dictionary (\ref{dic}) is supported also by another argument based
on expanding a Wilson loop, following the method developed in
\cite{Miwa}.

Let us consider the Taylor expansion of Wilson line
\[
 W(C) = {\rm Tr}\left[
{\rm P}\,\exp\left(\int^{s_f}_{s_i}\!\!ds\,\left(i A_{\mu}(x(s))\dot{x}^{\mu}(s) 
+ \phi_i\dot{y}^i(s) \right)\right)
\right]
\]
around the straight Wilson line $C_0$ 
\[
 x^{0}_{C_0}(s) = s\,, \qquad \dot{y}^i_{C_0}(s) = (0,0,0,0,0,1)\,, 
\]
by considering a small deformation like $W(C)=W(C_0 + \delta C)$
\cite{Miwa}. 

\medskip 

The Wilson line can be expanded as  
\begin{eqnarray}
W(C) &=& 
W(C_0) + \int^{s_f}_{s_i}\!ds\,\delta x^{\mu}(s)\left.\frac{\delta W(C)}{\delta
 x^{\mu}(s)}\right|_{C=C_0} + \int^{s_f}_{s_i}\!ds\,\delta\dot{y}^i(s)\left.
\frac{\delta W(C)}{\delta\dot{y}^i(s)}\right|_{C=C_0} \nonumber \\ 
&& \quad + \frac{1}{2}\int^{s_f}_{s_i}\!ds_1\!\int^{s_f}_{s_i}\!\!
ds_2\,\left.\delta
 x^{\mu}(s_1)\delta x^{\nu}(s_2)\frac{\delta^2 W(C)}{\delta
 x^{\mu(s_1)}\delta x^{\nu}(s_2)}\right|_{C=C_0} + \cdots\,.
\end{eqnarray}
The zeroth order term is nothing but the straight Wilson line. 
Then the first order term is evaluated as 
\begin{eqnarray}
\left.\frac{\delta W(C)}{\delta x^{\mu}(s)}\right|_{C=C_0} 
&=& {\rm Tr}\biggl[\left(
iF_{\mu\nu}(x(s))\dot{x}^{\nu}(s) + D_{\mu}\phi_i(x(s))\dot{y}^i(s)
\right) \times 
\nonumber \\ 
&& \left.\qquad \times {\rm P}\,\exp\left(\int^{s+s_f}_{s+s_i}\!ds'\,
\left(iA_{\mu}(x(s'))\dot{x}^{\mu}(s') 
+ \phi_i\dot{y}^i(s') \right)
\right)
\biggr]\right|_{C=C_0} \nonumber \\
&=&  
 {\rm Tr}\left[\left( iF_{\mu 0} + D_{\mu}\phi_6\right) 
{\rm P}\,\exp\left(\int^{\infty}_{-\infty}\!dt\,
\left(iA_{0} + \phi_6\right)\right)\right]\,, \label{4.4} \\ 
\left.\frac{\delta W(C)}{\delta \dot{y}^{i}(s)}\right|_{C=C_0} 
&=& {\rm Tr}\left[\,\phi_i ~{\rm P}\,\exp\left(\int^{\infty}_{-\infty}\!dt\,
 \left(iA_{0} + \phi_6 \right) \right)\right]\,. \label{4.5}
\end{eqnarray}
The fluctuations are expanded as 
\[
 \delta x^{\mu}(s) = \sum_{n=-\infty}^{\infty}\delta x^{\mu}_n 
{\rm e}^{2\pi i n s/(s_f-s_i)}\,, \qquad 
 \delta \dot{y}^{i}(s) = \sum_{n=-\infty}^{\infty}\delta \dot{y}^{i}_n 
{\rm e}^{2\pi i n s/(s_f-s_i)}\,,
\]
and those give the following relations:
\begin{eqnarray}
&& \sum_n \delta x_n^{\mu}\int^{s_f}_{s_i}\!ds\,\left.\frac{\delta W(C)}{\delta
 x^{\mu}(s)}\right|_{C=C_0} {\rm e}^{2\pi ins/(s_f-s_i)} 
= \delta x_0^{\mu} \times (\ref{4.4})\,, \nonumber \\ 
&& \sum_n \delta \dot{y}^i_n
\int^{s_f}_{s_i}\!ds\,\delta\dot{y}^i(s)\left.
\frac{\delta W(C)}{\delta\dot{y}^i(s)}\right|_{C=C_0}  
{\rm e}^{2\pi ins/(s_f-s_i)} 
= \delta \dot{y}_0^{i} \times (\ref{4.5})\,. \nonumber
\end{eqnarray}
By letting 
\begin{eqnarray}
\delta  x^0=\delta \dot y^6=0\,,  \label{cdd}
\end{eqnarray}
we are left with the impurities given in (\ref{dic}).
The condition (\ref{cdd}) will be supported from a supersymmetry 
argument.

\subsubsection*{Supersymmetry}

The straight Wilson line $W(C_0)$ is 1/2 BPS. As one can easily see, the
supersymmetry variation\footnote{ The supersymmetry
transformation is given by $\delta_\epsilon
A_\mu=i\bar\Psi\Gamma_\mu\epsilon$ and $\delta_\epsilon
\phi_i=i\bar\Psi\Gamma_i\epsilon$\,.} $\delta_\epsilon W(C_0)$ vanishes
when
\[
 (\dot x^\mu_{C_0}\Gamma_\mu-i\dot y^i_{C_0}\Gamma_i)\epsilon=0\,, 
\]
which implies  the locally supersymmetric condition
\[
(\dot x^\mu_{C_0})^2-(\dot y^i_{C_0})^2=0\,.
\]   
The straight Wilson line $W(C_0)$ indeed satisfies the condition. 

Then let us consider linear fluctuations around $W(C_0)$\,.   
The supersymmetry transformation for the fluctuations vanishes when 
the following relation is satisfied: 
\[
\left[(\dot x^\mu_{C_0}+\delta\dot x^\mu)\Gamma_\mu
-i(\dot y^i_{C_0}+\delta\dot y^i)\Gamma_i\right]\epsilon=0\,. 
\]
It implies that 
\[
\delta \dot x^0-\delta\dot y^6=0\,. 
\] 
On the other hand, one may choose as 
\[
\delta \dot x^0+\delta\dot y^6=0
\] 
by using the SO(1,1) symmetry. Thus the linear fluctuations are 
locally supersymmetric and 1/2 BPS when $\delta \dot x^0=0$ and
$\delta\dot y^6=0$\,. The former means $\delta x^0=0$ as the Wilson line
is characterized by $x^\mu$.
In fact, we have just seen
above that impurities (\ref{dic}) appeared as fluctuations satisfying
$\delta x^0=\delta \dot y^6=0$\,.

\subsection{Supergravity Modes in Non-relativistic Limit}

Next let us consider the fluctuations by focusing upon the supergravity 
modes. These should be BPS and protected from quantum corrections. 
Then the mass dimensions of the fluctuations
can be computed following \cite{AdS/CFT:GKPW}.

We show that
the mass dimensions of the fluctuations
propagating in \ads{2}
evaluated at the boundary
are equal to the conformal dimensions of the conjectured
impurities.
The masses of the fluctuations are
easily derived from
(\ref{fluctuation action:bosonic})
with $F=0$ and 
(\ref{fluctuation action:fermionic})
as
\begin{eqnarray}
m_B^2=2~~\mbox{for}~~\bar\phi~,~~~
m_B^2=0~~\mbox{for}~~\tilde\gamma,~\tilde\varphi \label{mass}
\end{eqnarray}
and $m_F^2=1$ for $\vartheta$.
The mass dimensions
of scalars can be evaluated at the boudary as follows \cite{AdS/CFT:GKPW}: 
\begin{eqnarray}
\Delta=\frac{1}{2}(1+\sqrt{1+4m^2})\,. \label{GKPW}
\end{eqnarray} 
By substituting (\ref{mass}) for (\ref{GKPW}), it follows that
\begin{eqnarray}
\Delta(\bar\phi)=2
~~~\mbox{and}~~~
\Delta(\tilde\gamma,\tilde\varphi)=1\,. \nonumber 
\end{eqnarray}
On the other hand, the conformal dimensions
of impurities are
\begin{eqnarray}
2
~~\mbox{for}~~
D_{a}\phi_6 + iF_{a0}
~~~\mbox{and}~~~
1
~~\mbox{for}~~
\phi^{a'}\,. \nonumber  
\end{eqnarray}
Thus we find the agreement of the mass dimensions
of the fluctuations and 
the conformal dimensions of the impurities.
This  provides a further support of our conjecture.

\medskip  

Finally we comment on the stringy excitation modes. The quantum spectrum
of the non-relativistic string has not been obtained yet by solving the
theory. The world-sheet theory is free but the world-sheet geometry is
\ads{2} rather than flat, and the quantization of it is more
involved. Furthermore, unlike the pp-wave string case, the U(1) charge
$J$ is not included in the present case.  Thus the BMN trick does not
work and hence perturbation theory would be useless in comparing the
gauge side with the string result. 
Anyway, it would be difficult
to check the above dictionary at stringy level.

\section{Summary and Discussions}

We have derived non-relativistic actions of string and D-branes on 
\adss{5}{5} from a semiclassical
approximation around static configurations. 
Then the AdS/CFT dictionary in the non-relativistic limit, which is a
new solvable sector pointed out in \cite{GGK}, has been argued on the
basis of the semiclassical interpretation and the symmetry argument. We
speculate that a state of non-relativistic string would correspond to a
small deformation of 1/2 BPS straight Wilson line. 

\medskip 

Here it may be valuable to comment on the difference of setups between
ours and \cite{DK,MY}. In the case of \cite{DK,MY}
a deformed Wilson loop with ``long'' composite operators is 
considered and large U(1)$_{\rm R}$ charges are included as the length
of the long operators unlike our case. Therefore
perturbation theory works in the case of \cite{DK,MY} to check
the correspondence even at stringy level and it is also an interesting
direction to study the deformed Wilson loops furthermore (For the works
in this direction, see \cite{TZ,DK,MY,Tsuji}).
On the other hand, it would be quite difficult for our case to test 
at the strigy level.  

\medskip 

It would be another direction to study a semiclassical approximation of
M-branes. The non-relativistic M-brane actions have already been
obtained in our previous paper \cite{SY:non-rela}.  We will report on
this issue in another place in the near future \cite{future}.

\medskip 

We hope that our result could be a new clue of the study of AdS/CFT
correspondence.

\section*{Acknowledgment}

We would like to thank Akitsugu Miwa, Joaquim Gomis, Machiko Hatsuda,
Yasuaki Hikida, Kiyoshi Kamimura, Tadashi Takayanagi for useful
discussions, and especially Satoshi Yamaguchi for intensive discussion
and comments at an early stage in this project. This work is supported
in part by the Grant-in-Aid for Scientific Research (No.~17540262 and
No.~17540091) from the Ministry of Education, Science and Culture,
Japan. The work of K.~Y.\ is supported in part by JSPS Research
Fellowships for Young Scientists.

\appendix

\section*{Appendix}

\section{The supervielbeins on the \adss{5}{5}} 

Here we will construct the supervielbeins on the \adss{5}{5}
background, in order to make the manuscript self-contained and 
fix the notations and conventions. 

First of all, let us introduce the metric of \adss{5}{5} background: 
\begin{eqnarray}
ds^2_{AdS}&=&-\cosh^2\rho dt^2
+d\rho^2
+\sinh^2\rho 
\Bigl[d\phi_1^2+\cos^2\phi_1(d\phi_2^2+\cos^2\phi_2d\phi_3^2)\Bigr]\,, 
\label{metric: AdS5}
\\
ds^2_{S}&=&
d\gamma^2
+\cos^2\gamma 
\Bigl[d\varphi_1^2+\cos^2\varphi_1 (
d\varphi_2^2+\cos^2\varphi_2
\bigl(d\varphi_3^2+\cos^2\varphi_3d\varphi_4^2)\bigr)
\Bigr]\,.
\label{metric: S5}
\end{eqnarray}
As we factor out the radii $R$ of \ads{5} and \s{5} to be an overall factor in the action,
the geometrical objects collected here are for $R=1$.
Then for the \ads{5}\, part, one can read off from (\ref{metric: AdS5})
the vielbein $e^a$ ($a=0,1,\cdots,4$) as
\begin{eqnarray}
e^a&=&(\cosh\rho dt,\,
d\rho,\,
\sinh\rho d\phi_1,\,
\sinh\rho\cos\phi_1d\phi_2,\,
\sinh\rho\cos\phi_1\cos\phi_2d\phi_3)\,. \nonumber 
\end{eqnarray}
The spin connection is defined by $de^a=-\omega^a{}_be^b$\,, and the
non-trivial components of the spin connection for the \ads{5} part, are
\begin{eqnarray}
&&
\omega^0{}_1=\sinh\rho dt\,, \quad 
\omega^2{}_1=\cosh\rho d\phi_1\,, \quad 
\omega^3{}_1=\cosh\rho\cos\phi_1d\phi_1\,, \quad 
\omega^3{}_2=-\sin\phi_1 d\phi_2\,,
\nonumber \\ &&
\omega^4{}_1=\cosh\rho\cos\phi_1\cos\phi_2 d\phi_3\,, \quad 
\omega^4{}_2=-\sin\phi_1\cos\phi_2 d\phi_3\,, \quad 
\omega^4{}_3=-\sin\phi_2 d\phi_3\,. \nonumber 
\end{eqnarray}
Next, for the S$^5$ part, the vielbein $e^{a'}$ ($a'=5,6,\cdots,9$) is seen
from (\ref{metric: S5}) as
\begin{eqnarray}
e^{a'} &=&(
d\gamma,\,
\cos\gamma d\varphi_1,\,
\cos\gamma \cos\varphi_1 d\varphi_2,\,
\nonumber\\&& \qquad 
\cos\gamma\cos\varphi_1\cos\varphi_2 d\varphi_3,\,
\cos\gamma\cos\varphi_1\cos\varphi_2\cos\varphi_3 d \varphi_4 )\,. 
\nonumber 
\end{eqnarray}
The non-zero components of the spin connection 
are given by 
\begin{eqnarray}
&&
\omega^6{}_5=-\sin\gamma d\varphi_1\,, \quad 
\omega^7{}_5=-\sin\gamma\cos\varphi_1 d\varphi_2\,, \quad 
\omega^7{}_6=-\sin\varphi_1d\varphi_2~,~~
\nonumber\\&&
\omega^8{}_5=-\sin\gamma\cos\varphi_1\cos\varphi_2 d\varphi_3\,, \quad 
\omega^8{}_6=-\sin\varphi_1\cos\varphi_2 d\varphi_3\,, \quad 
\omega^8{}_7=-\sin\varphi_2d\varphi_3\,, \quad 
\nonumber\\&&
\omega^9{}_5=-\sin\gamma\cos\varphi_1\cos\varphi_2\cos\varphi_3 
d\varphi_4\,, \quad 
\omega^9{}_6=-\sin\varphi_1\cos\varphi_2\cos\varphi_3 d\varphi_4\,, \quad 
\nonumber \\ &&
\omega^9{}_7=-\sin\varphi_2\cos\varphi_3 d\varphi_4\,, \quad 
\omega^9{}_7=-\sin\varphi_3 d\varphi_4\,. \nonumber 
\end{eqnarray}

\medskip 

Then it is turn to consider the supervielbeins on the \adss{5}{5}\,,
which can be obtained via the coset construction with the coset
superspace:
\begin{eqnarray}
\mbox{\adss{5}{5}} \sim \frac{PSU(2,2|4)}{SO(1,4)\times SO(5)}\,. \label{group}
\end{eqnarray} 
The super-\adss{5}{5} algebra is represented by the super Lie algebra 
$psu(2,2|4)$\,, whose commutation relations are given by 
\begin{eqnarray}
&&
{[}P_a,P_b]=
\alpha^2J_{ab}\,, \quad 
{[}P_{a'},P_{b'}]=
-\alpha^2J_{a'b'}\,, 
\nonumber\\&&
{[}P_a,J_{bc}]=\eta_{ab}P_c-\eta_{ac}P_b\,, \quad 
{[}P_{a'},J_{b'c'}]=\eta_{a'b'}P_{c'}-\eta_{a'c'}P_{b'}\,,
\nonumber\\&&
{[}J_{ab},J_{cd}]=\eta_{bc}J_{ad}+\text{3-terms}\,, \quad 
{[}J_{a'b'},J_{c'd'}]=\eta_{b'c'}J_{a'd'}+\text{3-terms}\,,
\nonumber\\&&
{[}Q_I,P_A]=
-\frac{\alpha}{2}Q_J(i\sigma_2)_{JI}\widehat\Gamma_{A}\,, \quad 
{[}Q_I,J_{AB}]=
-\frac{1}{2}Q_I\Gamma_{AB}\,,
\label{AdS5xS5 algebra} 
\\&&
\{Q_I,Q_J\}=
2i\CC\Gamma^A\delta_{IJ}h_+P_A
-i\alpha\CC\widehat\Gamma^{AB}(i\sigma_2)_{IJ}h_+J_{AB}\,, \nonumber 
\end{eqnarray}
where $A=(a,a')$.
$\alpha\equiv 1/R$ is set to be $1$ in the body of the paper.

Then the gamma matrix $\Gamma^A\in$ Spin(1,9) satisfies
\begin{eqnarray}
\{\Gamma^A,\Gamma^B\}=2\eta^{AB}~,~~~
(\Gamma^A)^T=-\CC\Gamma^A\CC^{-1}~,~~~
\CC^T=-\CC\,, \nonumber 
\end{eqnarray}
where $\CC$ is the charge conjugation matrix and the Minkowski metric
$\eta_{AB}$ is almost positive. Furthermore we have introduced the
following quantities: 
\begin{eqnarray}
&&\widehat\Gamma_A=(-\Gamma_a\CI,\Gamma_{a'}\CJ)~,~~~
\widehat\Gamma_{AB}=(-\Gamma_{ab}\CI,\Gamma_{a'b'}\CJ)~,~~~
\CI=\Gamma^{01234}~,~~~\CJ=\Gamma^{56789}~,
\nonumber\\&&
Q_Ih_+=Q_I\,, \quad 
h_+=\frac{1}{2}(1+\Gamma_{11})\,, \quad 
\Gamma_{11}=\Gamma_{01\cdots9}\,.  \nonumber 
\end{eqnarray}

\medskip 

Now the supervielbeins $\BL^{A}$ and $L^\alpha$, and super spin connection
$\BL^{AB}$ can be read from the left-invariant Cartan one-form defined by
\begin{eqnarray}
g^{-1}dg&=&\BL^AP_A +\frac{1}{2}\BL^{AB}J_{AB}
+Q_\alpha L^\alpha\,, \nonumber 
\end{eqnarray}
where $g$ is an element of the supercoset (\ref{group}) and it is
parametrized as
\begin{eqnarray}
g=g_xg_\theta\,, \quad 
g_{\theta}=\e^{Q\theta}\,, \quad Q=(Q_1,Q_2)\,, \quad 
\theta=\left(
  \begin{array}{c}
    \theta_1   \\
    \theta_2   \\
  \end{array}
\right)\,, \nonumber 
\end{eqnarray}
and $g_x$ and $g_{\theta}$ are the bosonic and the fermionic elements,
respectively. Here we note that $g_x$ should satisfy, by definition,
that
\begin{eqnarray}
g_x^{-1}dg_x&=&e^AP_A+\frac{1}{2}\omega^{AB}J_{AB}\,,  \nonumber 
\end{eqnarray}
where $e^A$ and $\omega^{AB}$ are the vielbein and the spin connection
of the \adss{5}{5}\,. 

After some algebra, we finally obtain the explicit expressions of the
supervielbeins and super spin connection as follows\footnote{ The
differential $d$ acts as $d(F\wedge G)=dF\wedge G+(-1)^{f}F\wedge dG$
(where $f$ is the degree of $F$), and commutes with $\theta$.  }
\begin{eqnarray}
\BL^A&=&
e^A+2i\sum_{n=1}^\infty\bar\theta\Gamma^A\frac{\CM^{2n-2}}{(2n)!}D\theta
=
e^A+2i\bar\theta\Gamma^A
\left(
\frac{\cosh\CM-1}{\CM^2}
\right)
D\theta\,, \nonumber \\
L^\alpha&=&
\sum_{n=0}^\infty\frac{\CM^{2n}}{(2n+1)!}D\theta
=
\frac{\sinh\CM}{\CM}D\theta\,, \nonumber \\
\BL^{AB}&=&
\omega^{AB}
-2i\alpha\bar\theta\widehat\Gamma^{AB}i\sigma_2
\sum_{n=1}^\infty\frac{\CM^{2n-2}}{(2n)!}D\theta
=\omega^{AB}
-2i\alpha\bar\theta\widehat\Gamma^{AB}i\sigma_2
\frac{\cosh\CM-1}{\CM^2}
D\theta \nonumber 
\end{eqnarray}
with
\begin{eqnarray}
\CM^2&=&
i\alpha\left(
\widehat\Gamma_Ai\sigma_2\theta\,
\bar\theta\Gamma^A
-\frac{1}{2}\Gamma_{AB}\theta\,
\bar\theta\widehat\Gamma^{AB}i\sigma_2
\right)\,, \nonumber \\
D\theta&=&
d\theta
+\frac{\alpha}{2}e^A\widehat\Gamma_Ai\sigma_2\theta
+\frac{1}{4}\omega^{AB}\Gamma_{AB}\theta\,.
\end{eqnarray}
By construction these should satisfy the following MC equations: 
\begin{eqnarray}
d\BL^A&=&
-\eta_{BC}\BL^{AB}\BL^C
+i\bar L\Gamma^AL~,\nonumber\\
d\BL^{ab}&=&
-\alpha^2\BL^a\BL^b
-\eta_{cd}\BL^{ca}\BL^{bd}
-i\alpha\bar L\widehat\Gamma^{ab} i\sigma_2 L~,
\label{MC AdSxS in 10-dim} \\
d\BL^{a'b'}&=&
+\alpha^2\BL^{a'}\BL^{b'}
-\eta_{c'd'}\BL^{c'a'}\BL^{b'd'}
-i\alpha\bar L\widehat\Gamma^{a'b'} i\sigma_2 L~,\nonumber\\
dL^\alpha&=&
-\frac{\alpha}{2}\BL^A\widehat\Gamma_A i\sigma_2L
-\frac{1}{4}\BL^{AB}\Gamma_{AB}L\,, \nonumber 
\end{eqnarray}
which are equivalent to the $psu(2,2|4)$ algebra (\ref{AdS5xS5 algebra}).


\begin{thebibliography}{99}



\bibitem{AdS/CFT:M}
J.~M.~Maldacena,
``The large N limit of superconformal field theories and supergravity,''
Adv.\ Theor.\ Math.\ Phys.\  {\bf 2} (1998) 231
[Int.\ J.\ Theor.\ Phys.\  {\bf 38} (1999) 1113] [arXiv:hep-th/9711200].

\bibitem{AdS/CFT:GKPW}
S.~S.~Gubser, I.~R.~Klebanov and A.~M.~Polyakov,
``Gauge theory correlators from non-critical string theory,''
Phys.\ Lett.\ B {\bf 428} (1998) 105 [arXiv:hep-th/9802109];
E.~Witten, ``Anti-de Sitter space and holography,''
Adv.\ Theor.\ Math.\ Phys.\  {\bf 2} (1998) 253 [arXiv:hep-th/9802150]. 


\bibitem{MT}
  R.~R.~Metsaev and A.~A.~Tseytlin,
  ``Type IIB superstring action in \adss{5}{5} background,''
  Nucl.\ Phys.\ B {\bf 533} (1998) 109
  [arXiv:hep-th/9805028].

\bibitem{BPR}
  I.~Bena, J.~Polchinski and R.~Roiban,
  ``Hidden symmetries of the \adss{5}{5} superstring,''
  Phys.\ Rev.\ D {\bf 69} (2004) 046002
  [arXiv:hep-th/0305116].

\bibitem{BMN}
  D.~Berenstein, J.~M.~Maldacena and H.~Nastase,
   ``Strings in flat space and pp waves from 
$\mathcal{N}$=4 super Yang Mills,''
  JHEP {\bf 0204}, 013 (2002)
  [arXiv:hep-th/0202021].
\bibitem{Penrose}
R.~Penrose,
``Any spacetime has a plane wave as a limit,''
Differential geometry and relativity, Reidel, Dordrecht, 1976,
pp.~271-275; \\
R.~Gueven,
``Plane wave limits and T-duality,''
Phys.\ Lett.\ B {\bf 482} (2000) 255
[arXiv:hep-th/0005061]; \\
M.~Blau, J.~Figueroa-O'Farrill, C.~Hull and
G.~Papadopoulos, ``Penrose limits and maximal supersymmetry,'' Class.\
Quant.\ Grav.\ {\bf 19} (2002) L87 [arXiv:hep-th/0201081]. 

\bibitem{Metsaev}
  R.~R.~Metsaev,
  ``Type IIB Green-Schwarz superstring in plane wave Ramond-Ramond
  background,''
  Nucl.\ Phys.\ B {\bf 625} (2002) 70
  [arXiv:hep-th/0112044]. 

\bibitem{MT2}
  R.~R.~Metsaev and A.~A.~Tseytlin,
  ``Exactly solvable model of superstring in plane wave Ramond-Ramond
  background,''
  Phys.\ Rev.\ D {\bf 65} (2002) 126004
  [arXiv:hep-th/0202109].

\bibitem{GGK} J.~Gomis, J.~Gomis and K.~Kamimura, 
``Non-relativistic superstrings: A new soluble sector 
of \adss{5}{5},'' 
JHEP {\bf 0512} (2005) 024 [arXiv:hep-th/0507036].

\bibitem{GO}
  J.~Gomis and H.~Ooguri,
   ``Non-relativistic closed string theory,''
  J.\ Math.\ Phys.\  {\bf 42}, 3127 (2001)
  [arXiv:hep-th/0009181]. 


\bibitem{Brugues:2004an}
  J.~Brugues, T.~Curtright, J.~Gomis and L.~Mezincescu,
   ``Non-relativistic strings and branes as non-linear realizations of  Galilei
   groups,''
  %
  Phys.\ Lett.\ B {\bf 594}, 227 (2004)
  [arXiv:hep-th/0404175]
\bibitem{Gomis:2004pw}
  J.~Gomis, K.~Kamimura and P.~K.~Townsend,
   ``Non-relativistic superbranes,''
  %
  JHEP {\bf 0411}, 051 (2004)
  [arXiv:hep-th/0409219].
\bibitem{Gomis:2005bj}
  J.~Gomis, F.~Passerini, T.~Ramirez and A.~Van Proeyen,
  ``Non relativistic D$p$-branes,''
  JHEP {\bf 0510} (2005) 007
  [arXiv:hep-th/0507135].
\bibitem{Kamimura:2005rz}
K.~Kamimura and T.~Ramirez,
  ``Brane dualities in non-relativistic limit,''
  JHEP {\bf 0603} (2006) 058
  [arXiv:hep-th/0512146].
\bibitem{Brugues:2006yd}
  J.~Brugues, J.~Gomis and K.~Kamimura,
   ``Newton-Hooke algebras, non-relativistic branes and generalized pp-wave
  metrics,''
  Phys.\ Rev.\ D {\bf 73} (2006) 085011
  [arXiv:hep-th/0603023].

\bibitem{GP}
  G.~W.~Gibbons and C.~E.~Patricot,
  ``Newton-Hooke space-times, Hpp-waves and the cosmological constant,''
  Class.\ Quant.\ Grav.\  {\bf 20} (2003) 5225
  [arXiv:hep-th/0308200].





\bibitem{GKP}
  S.~S.~Gubser, I.~R.~Klebanov and A.~M.~Polyakov,
  ``A semi-classical limit of the gauge/string correspondence,''
  Nucl.\ Phys.\ B {\bf 636} (2002) 99
  [arXiv:hep-th/0204051].
  
  
\bibitem{Frolov:2002av}
  S.~Frolov and A.~A.~Tseytlin,
  ``Semiclassical quantization of rotating superstring in \adss{5}{5},''
  JHEP {\bf 0206} (2002) 007
  [arXiv:hep-th/0204226].
  
\bibitem{SY:non-rela}
  M.~Sakaguchi and K.~Yoshida,
  ``Non-relativistic AdS branes and Newton-Hooke superalgebra,''
  JHEP {\bf 0610} (2006) 078
  [arXiv:hep-th/0605124].




  \bibitem{DGT}
  N.~Drukker, D.~J.~Gross and A.~A.~Tseytlin,
  ``Green-Schwarz string in \adss{5}{5}: Semiclassical partition  function,''
  JHEP {\bf 0004} (2000) 021
  [arXiv:hep-th/0001204].

\bibitem{Wilson}
  S.~J.~Rey and J.~T.~Yee,
  ``Macroscopic strings as heavy quarks in large N gauge theory and  anti-de
  Sitter supergravity,''
  Eur.\ Phys.\ J.\ C {\bf 22} (2001) 379
  [arXiv:hep-th/9803001]. \\ 
  J.~M.~Maldacena,
  ``Wilson loops in large N field theories,''
  Phys.\ Rev.\ Lett.\  {\bf 80} (1998) 4859
  [arXiv:hep-th/9803002].



\bibitem{D:curved} 
M.~Cederwall, A.~von Gussich, B.~E.~W.~Nilsson, P.~Sundell 
and A.~Westerberg, 
``The Dirichlet super-p-branes in ten-dimensional type IIA and 
IIB supergravity,'' 
Nucl.\ Phys.\ B {\bf 490} (1997) 179 [arXiv:hep-th/9611159];\\ 
E.~Bergshoeff and P.~K.~Townsend, 
``Super D-branes,'' 
Nucl.\ Phys.\ B {\bf 490} (1997) 145 [arXiv:hep-th/9611173].

\bibitem{D:flat}
  M.~Aganagic, C.~Popescu and J.~H.~Schwarz,
  ``D-brane actions with local kappa symmetry,''
  Phys.\ Lett.\ B {\bf 393} (1997) 311
  [arXiv:hep-th/9610249];\\
  ``Gauge-invariant and gauge-fixed D-brane actions,''
  Nucl.\ Phys.\ B {\bf 495} (1997) 99
  [arXiv:hep-th/9612080].


\bibitem{SMT}
K.~Skenderis and M.~Taylor,
``Branes in AdS and pp-wave spacetimes,''
JHEP {\bf 0206} (2002) 025 [arXive:hep-th/0204054]. 

\bibitem{Sakaguchi:2003py} M.~Sakaguchi and K.~Yoshida, ``D-branes of
covariant AdS superstrings,'' Nucl.\ Phys.\ B {\bf 684} (2004) 100
[arXiv:hep-th/0310228]; \\ For a proof in the full order of $\theta$\,,
see ``Notes on D-branes of type IIB string on \adss{5}{5},''
Phys.\ Lett.\ B {\bf 591} (2004) 318 [arXiv:hep-th/0403243]; \\ For a
short summary, see  ``D-branes of covariant AdS superstrings: An overview,''
arXiv:hep-th/0408208. \\ 
For a generalization including gauge field condensates, see 
  ``Noncommutative D-brane from covariant AdS superstring,''
  arXiv:hep-th/0604039.

\bibitem{D3:AdS}
  R.~R.~Metsaev and A.~A.~Tseytlin,
  ``Supersymmetric D3 brane action in \adss{5}{5},''
  Phys.\ Lett.\ B {\bf 436} (1998) 281
  [arXiv:hep-th/9806095].

\bibitem{D3:pp}
  R.~R.~Metsaev,
  ``Supersymmetric D3 brane and N = 4 SYM actions in plane wave backgrounds,''
  Nucl.\ Phys.\ B {\bf 655} (2003) 3
  [arXiv:hep-th/0211178].


\bibitem{giant}
  J.~McGreevy, L.~Susskind and N.~Toumbas,
  ``Invasion of the giant gravitons from anti-de Sitter space,''
  JHEP {\bf 0006} (2000) 008
  [arXiv:hep-th/0003075]. \\ 
  M.~T.~Grisaru, R.~C.~Myers and O.~Tafjord,
  ``SUSY and Goliath,''
  JHEP {\bf 0008} (2000) 040
  [arXiv:hep-th/0008015]. \\ 
  A.~Hashimoto, S.~Hirano and N.~Itzhaki,
  ``Large branes in AdS and their field theory dual,''
  JHEP {\bf 0008} (2000) 051
  [arXiv:hep-th/0008016].

\bibitem{gWilson}
  J.~Gomis and F.~Passerini,
  ``Holographic Wilson loops,''
  JHEP {\bf 0608} (2006) 074
  [arXiv:hep-th/0604007].


\bibitem{Miwa}
  A.~Miwa,
  ``BMN operators from Wilson loop,''
  JHEP {\bf 0506} (2005) 050
  [arXiv:hep-th/0504039].


\bibitem{DK}
  N.~Drukker and S.~Kawamoto,
  ``Small deformations of supersymmetric Wilson loops and open spin-chains,''
  JHEP {\bf 0607} (2006) 024
  [arXiv:hep-th/0604124].

\bibitem{MY}
  A.~Miwa and T.~Yoneya,
  ``Holography of Wilson-loop expectation values with local operator
  insertions,''
  arXiv:hep-th/0609007.

\bibitem{TZ}
  A.~A.~Tseytlin and K.~Zarembo,
  ``Wilson loops in $\mathcal{N}$=4 SYM theory: Rotation in S$^5$,''
  Phys.\ Rev.\ D {\bf 66} (2002) 125010
  [arXiv:hep-th/0207241].

\bibitem{Tsuji}
  A.~Tsuji,
  ``Holography of Wilson loop correlator and spinning strings,''
  arXiv:hep-th/0606030. 

\bibitem{future}
M.~Sakaguchi and K.~Yoshida, in preparation.


\end{thebibliography}
\end{document}